\documentstyle[amssymb,preprint,aps]{revtex}

\begin{document}
\title{Coulomb-enhanced dynamic localization and Bell state generation in coupled
quantum dots}
\author{Ping Zhang$^{1,2}$, Qi-Kun Xue$^{1}$, Xian-Geng Zhao$^{2}$, X.C. Xie$^{3,1}$}
\address{$^1$International Center of Quantum Structure and State Key\\
laboratory for\\
Surface Physics, Institute of Physics, The Chinese Academy of Sciences,\\
Beijing 100080, P.R. China\\
$^2$Institute of Applied Physics and Computational Mathematics, Beijing\\
100080, P.R. China\\
$^3$Department of Physics, Oklahoma State University, Stillwater, OK 74078 }
\maketitle

\begin{abstract}
We investigate the dynamics of two interacting electrons in coupled quantum
dots driven by an AC field. We find that the two electrons can be trapped in
one of the dots by the AC field, in spite of the strong Coulomb repulsion.
In particular, we find that the interaction may enhance the localization
effect. We also demonstrate the field excitation procedure to generate the
maximally entangled Bell states. The generation time is determined by both
analytic and numerical solutions of the time dependent Schr\"{o}dinger
equation.%
\newline%
\newline%
\end{abstract}

\section{Introduction}

Quantum-state engineering via optical or electrical manipulation over the
coherent dynamics of suitable quantum-mechanical systems has become a
fascinating prospect of modern physics. A very intriguing result in the
study of a single particle in a double-trap system exposed to a
time-dependent external field is the dynamic localization phenomenon\cite
{Grossman,Metiu,Hanggi}, i.e., for appropriate field parameters, a localized
wave packet remains dynamically localized during the subsequent time
evolution. This driving induced trapping has been theoretically and
experimentally studied in many physical and chemical systems\cite
{Hol,Zuo,Raizen}.

When two or more interacting particles are present, apart from the highly
nontrivial question of whether the strong many-body interaction can be
overcome for the particles to create and preserve localization, the
possibility of entanglement of the many-body wave functions arises.
Entanglement is an essential ingredient in any scheme of quantum information
processing. Recently, quantum dot realizations of the entanglement have
received increasing attention. Various schemes based on electron spins and
electron-hole pairs have been proposed to implement quantum computer
hardware architectures\cite{Loss,Imamoglu,Johnson,Eli,Burcard}. Although
there have been some numerical studies on interacting electron systems
driven by an AC field\cite{Ping2,Tam2,Creff}, there is still little
theoretical understanding of the observed effects beyond the phenomenology
level at the present time.

In this paper, we address the dynamic localization and quantum entanglement
of two interacting electrons in a double quantum dot system driven by an AC
field (see Fig. 1). We show that with certain choice of parameters, in
contrast to direct intuition, the Coulomb repulsion may enhance the
localization due to the level crossing associated with different symmetries
of a dynamic parity operation (discussed in more detail later). We also show
that the maximally entangled Bell state can be prepared and maintained with
a pulse of an AC field. Our study provides useful information for the future
exploitation of coherent control of two-electron states in quantum dots.

In Sec. II we present the Hamiltonian for two interacting electrons in a
double-dot system. A simplified spin-1 representation of the Hamiltonian is
given in Sec. III. In Sec. IV we discuss the phenomenon of Coulomb-enhanced
localization. The field excitation procedure to generate the maximally
entangled Bell state is showed in Sec. V. A summary is given in Sec. VI. \ 

\section{Theoretical model}

The Hamiltonian which we use to describe the dynamics of two interacting
electrons in a coupled quantum dot driven by electric field is

\begin{equation}
H(t)=\sum_{i=1,2}h({\bf r}_{i},{\bf p}_{i},t)+V_{C}(|{\bf r}_{1}-{\bf r}%
_{2}|),  \eqnum{1a}
\end{equation}
\begin{equation}
h({\bf r},{\bf p},t)=\frac{{\bf p}^{2}}{2m^{\ast }}-ezF(t)+V_{t}({\bf r}%
)+V_{l}({\bf r}),  \eqnum{1b}
\end{equation}
\begin{equation}
V_{C}(|{\bf r}_{1}-{\bf r}_{2}|)=\frac{e^{2}}{\kappa |{\bf r}_{1}-{\bf r}%
_{2}|},  \eqnum{1c}
\end{equation}
where $V_{C}(|{\bf r}_{1}-{\bf r}_{2}|)$ is the Coulomb interaction and $h$
the single-particle Hamiltonian. The time-dependent electric field $F(t)$ is
applied along the inter-dot axis. The dielectric constant $\kappa $ and the
effective mass $m^{\ast }$ are material parameters. The potential $V_{t}$ in 
$h$ describes the in-plane confinement, whereas $V_{l}$ models the
longitudinal double-well structure. The transverse coupling of the dots is
modeled by a harmonic potential 
\begin{equation}
V_{t}(x,y)=\frac{m^{\ast }\omega _{t}^{2}}{2}(x^{2}+y^{2}).  \eqnum{2}
\end{equation}
It has been shown in experiments with electrically gated quantum dots in a
two-dimensional electron system that the electronic spectrum is well
described by a simple harmonic oscillator\cite{Jacak}. In describing the
confinement $V_{l}$ along the inter-dot axis, we use a (locally harmonic)
double well potential of the form

\begin{equation}
V_{l}=\frac{m^{\ast }\omega _{l}^{2}}{8a^{2}}(z^{2}-a^{2})^{2},  \eqnum{3}
\end{equation}
which, in the limit of large inter-dot distance, separates into two harmonic
wells (one for each dot) of frequency $\omega _{l}$. Although in principle a
square-well potential would be a more accurate description of the real
potential than the harmonic double well, there is no qualitative difference
between the results presented below obtained with harmonic potential and the
corresponding results using square-well potential.

We assume the transverse confinement is strong enough. Thus, the $(x,y)$
degrees of freedom are frozen in the dynamics and the two-electron wave
function can be written as $|\Psi ({\bf r}_{1},{\bf r}_{2},t)\rangle =|\phi
(x_{1},y_{1})\rangle |\phi (x_{2},y_{2})\rangle |\Phi (z_{1},z_{2},t)\rangle 
$, where $|\phi (x,y)\rangle $ is the transverse ground state. Considering
the fact that the external electric field is applied along the longitudinal
direction, the approximation of frozen in-plane motion is reasonable and has
been employed in previous work\cite{Tam2}. After integrating over the $(x,y)$
degrees of freedom, the time dependent Schr\"{o}dinger equation becomes 
\begin{eqnarray}
i\hslash \frac{\partial |\Phi (z_{1},z_{2},t)\rangle }{\partial t} &=&[-%
\frac{\hslash ^{2}}{2m^{\ast }}\left( \frac{\partial ^{2}}{\partial z_{1}^{2}%
}+\frac{\partial ^{2}}{\partial z_{2}^{2}}\right) +V(z_{1})+V(z_{2}) 
\eqnum{4} \\
&&+V_{C}(|z_{1}-z_{2}|)-eF(t)(z_{1}+z_{2})]|\Phi (z_{1},z_{2},t)\rangle , 
\nonumber
\end{eqnarray}
where $V_{C}$ is the effective one-dimensional Coulomb interaction 
\begin{equation}
V_{C}(|z_{1}-z_{2}|)=\int dx_{1}dx_{2}dy_{1}dy_{2}\frac{e^{2}\phi
^{2}(x_{1},y_{1})\phi ^{2}(x_{2},y_{2})}{\kappa |{\bf r}_{1}-{\bf r}_{2}|}. 
\eqnum{5}
\end{equation}
We use the effective mass $m^{\ast }$ and dielectric constant $\kappa $ of
GaAs. The other parameters are chosen as $\hslash \omega _{l}=20$meV and $%
a=20$nm.

The time evolution of the two electrons is obtained by numerically solving
the Schr\"{o}dinger equation (4) using an extension of the Crank-Nicholson
method\cite{Press} to two spatial dimensions. The initial state used in this
investigation is the field-free ground state, which is obtained by
propagating a trial two-particle wave packet in the imaginary time domain
and shown in Fig. 2(a). As illustrated by the spatial symmetry under the
particle exchange, the ground state is a singlet, as expected. Since Eq. (4)
contains no mixing between the singlet and triplet sub-spaces, the dynamics
of the system will always be confined to the singlet sub-space.

\section{Spin-1 representation of the Hamiltonian}

To understand the underlying physics behind the numerical results we employ
the Hund-Mulliken (HM) approximation by introducing the orthonormalized
one-particle wave functions $|\Phi _{\pm a}\rangle =(|\varphi _{\pm
a}\rangle -g|\varphi _{\mp a}\rangle )/\sqrt{1-2Sg+g^{2}}$, where $|\varphi
_{\pm a}\rangle $ are the single-particle ground states for the right and
left dots, $S=\langle \varphi _{+a}|\varphi _{-a}\rangle $ denotes the
overlap integral, and $g=(1-\sqrt{1-S^{2}})/S$. Using $|\Phi _{\pm a}\rangle 
$, we construct three singlet basis functions with respect to which we
diagonalize the two-electron Hamiltonian: Two states with double occupation
in each dot, $|RR\rangle =|\Phi _{+a}(z_{1})\rangle |\Phi
_{+a}(z_{2})\rangle $, $|LL\rangle =|\Phi _{-a}(z_{1})\rangle |\Phi
_{-a}(z_{2})\rangle $, and one state with single occupation in each dot, $%
|LR\rangle =(1/\sqrt{2})[|\Phi _{+a}(z_{1})\rangle |\Phi _{-a}(z_{2})\rangle
+|\Phi _{+a}(z_{2})\rangle |\Phi _{-a}(z_{1})\rangle ]$. Calculating the
matrix elements of the Hamiltonian in this orthonormal basis, we obtain 
\begin{equation}
H_{HM}(t)=uJ_{z}^{2}+2wJ_{x}-\mu (t)J_{z},  \eqnum{6}
\end{equation}
where we have dropped a constant energy term that makes no contribution to
the dynamics. In Eq. (6) $u=\langle LL|V_{C}|LL\rangle -\langle
LR|V_{C}|LR\rangle $ is the difference between the intradot and interdot
Coulomb interaction, $w=\langle \Phi _{\pm a}|h(z\mp a)|\Phi _{\mp a}\rangle
+\langle LR|V_{C}|LL\rangle $ denotes the single-particle tunneling induced
by dot-dot coupling and the Coulomb interaction, $\mu
(t)=2eF(t)a(1-g^{2})/(1-2Sg+g^{2})$ describes the electron-field coupling,
and $J_{x}$ and $J_{z}$ are the $x$- and $z$- components of the spin-1
operator. The localized two-particle state $|LL\rangle $ is equivalent to
the eigenstate $\mid j_{z}=1\rangle $ of $J_{z}$ and $|RR\rangle $ to the
state $\mid j_{z}=-1\rangle $, while the delocalized state $|LR\rangle $ is
identical to the state $\mid j_{z}=0\rangle $. According to the expression
for $u$ and $w$, using the material parameters given below Eq. (5) we obtain 
$u=12$meV and $w=-0.4$meV.

In the absence of an external driving, the eigenenergies of $H_{HM}$ are $%
E_{1}=(u-\sqrt{u^{2}+16w^{2}})/2,$ $E_{2}=u$, and $E_{3}=(u+\sqrt{%
u^{2}+16w^{2}})/2$, and the corresponding eigenstates are 
\begin{equation}
|\varphi _{1}\rangle =|LL\rangle -E_{3}/(\sqrt{2}w)|LR\rangle +|RR\rangle , 
\eqnum{7a}
\end{equation}
\begin{equation}
|\varphi _{2}\rangle =(|RR\rangle -|LL\rangle )/\sqrt{2},  \eqnum{7b}
\end{equation}
\begin{equation}
|\varphi _{3}\rangle =|LL\rangle -E_{1}/(\sqrt{2}w)|LR\rangle +|RR\rangle .%
\text{ \ \ }  \eqnum{7c}
\end{equation}
The symmetric ground state $|\varphi _{1}\rangle $ is dominated by the
delocalized state $|LR\rangle $ due to the strong Coulomb repulsion, whereas
the other two eigenstates are nearly degenerate and dominated by the two
localized states $|LL\rangle $ and $|RR\rangle $. The superposition of the
two localized states $|LL\rangle $ and $|RR\rangle $ implies that the
spatial wave functions of the two electrons have been entangled and
correlated, in the usual sense that they cannot be factorized into
single-particle states. The nonlinear term in $H_{HM}$ can be exploited to
generate entangled states. The ground state of Eq. (7) is plotted in Fig.
2(b). Clearly the HM approximation describes the ground state very well when
compared with the exact numerical solution [Fig. 2(a)].

\section{Coulomb-enhanced Dynamic localization}

To investigate dynamic localization, one must start with a localized wave
packet. This can be realized from the unperturbed ground state by two
separate methods: One is to suddenly switch on a DC field with the strength
satisfying the resonance condition $\mu _{0}=u$. At time $t\approx \pi /(%
\sqrt{2}w)$, the delocalized ground state $|\varphi _{1}\rangle $ evolves
into a localized state $|RR\rangle $ with two electrons occupying the same
right dot. The alternative method is to adiabatically switch on a constant
electric field. Then as shown in Fig. 2(c), the ground state configuration
develops a series of Coulomb stairs as a function of the field amplitude,
corresponding to a series of avoided crossing in the energy spectrum [see
Fig. 2(d)].

After preparing a localized state, say $|RR\rangle $, by adiabatic evolution
in the presence of a DC field, the DC field is suddenly swiched off and an
AC field of the form $F_{1}\sin (\omega t)$ is swiched on. We show now that
the localization can be dynamically maintained by the AC field even when the
field strength is small compared to the Coulomb repulsion between the two
electrons. Time periodicity of the Hamiltonian enables us to describe the
dynamics within the Floquet formalism. In addition, since the Hamiltonian is
invariant under the combined dynamic parity operation $z\rightarrow -z$; $%
t\rightarrow t+\pi /\omega $, each Floquet state is either odd or even.
Quasienergies of different parity may cross, otherwise an avoided crossing
may occur. Figures 3(a)-(b) show the quasienergy spectra of $H_{HM}(t)$ as a
function of $F_{1}$ with the presence and absence of the Coulomb
interaction, respectively. Two prominent features can be identified from the
comparison of these two cases: (i) the strong Coulomb interaction removes
the level crossings among three two-particle states and thus induces avoided
crossings; (ii) in the weak field regime, there occurs a crossing between
the quasienergies $\varepsilon _{2}$ and $\varepsilon _{3}$, which develop
from the unperturbed eigenenergies $E_{2}$ and $E_{3}$. To illustrate the
effect of this crossing on the system's dynamics, we begin with the initial
state $|RR\rangle $ and follow the time evolution of the probability $%
P_{RR}(t)$ for finding the two electrons in the right dot. The result is
shown in Fig. 3(c). It is clear that $P_{RR}$ always remains near 1 as if
the two electrons were frozen in the same right dot. This dynamic
localization seems counterintuitive at first as the Coulomb repulsion is
very strong compared to the AC field, preventing the two electrons occupying
the {\it same} dot. The localization shown in Fig. 3(c) has no
correspondence in a non-interacting case, since no level crossings occur in
the weak field regime if the Coulomb interaction is absent. Therefore, the
present localization effect results from an interplay of the Coulomb
interaction and an AC field. Furthermore, we find that even if the Coulomb
interaction is strong enough, the dynamic localization can still occur at
the crossing of the quasienergies $\varepsilon _{2}$ and $\varepsilon _{3}$.
Slightly away from the crossing point, the life time for finding two
electrons in the right dot is very long. This makes the experimental
realization of the dynamic localization more realistic. The ratio $2\mu
_{1}/\omega $\ at the first crossing is about $2.4$, a\ root of the
zero-order Bessel function, suggesting that the two-electron localization
can be approximated by a two-state model composed of $|\varphi _{2}\rangle $
and $|\varphi _{3}\rangle $.

Surprisingly, we find that in the strong field regime, the presence of
Coulomb repulsion may help to enhance the dynamic localization when compared
with the non-interacting case. From Eq. (7) we see that due to the strong
Coulomb interaction, $|RR\rangle \simeq (1/\sqrt{2})(|\varphi _{2}\rangle
+|\varphi _{3}\rangle )$. Thus the initial localized state can be
approximated by a superposition of the \ degenerate Floquet states developed
from the nearly-degenerate states $|\varphi _{2}\rangle $ and $|\varphi
_{3}\rangle $, which leads to a complete suppression of tunneling at the
crossing of the quasienergies $\varepsilon _{2}$ and $\varepsilon _{3}$ even
in the weak field limit. In the absence of the Coulomb interaction, the
initial localized state $|RR\rangle $ is a superposition of all three
eigenstates, suggesting dynamic localization at the crossing among three
quasienergies. However, the fundamental difference lies in the fact that
when the strong interaction is present, the tunneling coupling is $\langle
LL|H_{HM}|RR\rangle \simeq 4w^{2}/u$, whereas in the absence of the Coulomb
interaction the tunneling coupling is $2w$. Thus the coupling between the
two localized states greatly decreases in the presence of strong Coulomb
interaction, which leads to the enhancement of localization effects.

We emphasize that the localization only occurs at the crossing of the
quasienergies developed from the high-energy eigenstates $|\varphi
_{2}\rangle $ and $|\varphi _{3}\rangle $, in which case the delocalized
ground state $|\varphi _{1}\rangle $ is a dark state during time evolution.
For comparison, Figure 3(d) shows $P_{RR}(t)$ with the field parameters
corresponding to the crossing of the quasienergies $\varepsilon _{1}$ and $%
\varepsilon _{3}$. It reveals that dynamic localization disppears and the
two electrons start to oscillate between two dots. This fact highlights the
essential difference between a single-particle system and an interacting
two-particle system. The level crossing of $\varepsilon _{1}$\ and $%
\varepsilon _{3}$ induces strong participation of the Floquet state evolved
from the unperturbed ground state. As its largest component is the
delocalized two-particle state $|LR\rangle $, dynamic localization will be
completely destructed, as shown in Fig. 3(d).

\section{Bell state generation}

In this section we investigate the field excitation procedure to obtain a
maximally entangled Bell state of the form $|\Psi _{Bell}\rangle
=(|RR\rangle +e^{i\phi }|LL\rangle )$ with an arbitrary phase angle $\phi $.
Figure 4(a) plots the Floquet quasienergies of $H_{HM}(t)$ as a function of
the driving frequency $\omega $. It shows that when $\omega =u$, an avoided
crossing is formed between the quasienergies $\varepsilon _{1}$ and $%
\varepsilon _{2}$. Choosing the field parameters at this avoided crossing
and starting with the unperturbed ground state, we present in Fig. 4(b) the
probabilities $P_{LR}(t)$ for finding the two electrons in different dots, $%
P_{LL}$ in the left dot, and $P_{RR}$ in the right dot. It shows that the
two electrons oscillate between the delocalized state $|LR\rangle $ and two
localized states $|LL\rangle $ and $|RR\rangle $. The occupations of two
localized states are always the same and the oscillations are always
in-phase. Figure 4(c) shows the probability $\rho _{Bell}$ for finding the
maximally entangled Bell state with $\phi =\pi $. One can see that $\rho
_{Bell}$ varies with time, reaching a maximum value of 1 when $P_{LR}=0$
(the two electrons are maximally entangled with $\rho _{Bell}=1$). A direct
numerical solution of Eq. (4) gives the same prediction.

Once the two electrons are in the maximally entangled Bell state, they can
remain maximally entangled by suddenly turning off the AC field. We show
this effect in Figure 5. Figure 5(a) plots the time evolution of the
occupations of three two-particle states, and Figure 5(b) the probability $%
\rho _{Bell}$ with $\phi =\pi $ (solid line). It is clear from Fig. 5(a)
that a pulse of an AC field induces the two electrons to stay in the same
dot, while each of them occupies either of the dots with the same
probability. As shown in Fig. 5(b) (solid line), the two electrons remain
maximally entangled after the electric field is turned off. Therefore, the
maximally entangled Bell state can be created and maintained by applying a
pulse of a resonant AC field.

We derive an approximate analytical solution to highlight the physical
aspect of the Bell-state generation procedure. Taking into account symmetric
properties of the three unperturbed eigenstates of the system, the dynamics
shown in Fig. 5 is determined by the one-photon transition between the
ground state $|\varphi _{1}\rangle $ and the first excited state $|\varphi
_{2}\rangle $, whereas the transition from $|\varphi _{1}\rangle $ to $%
|\varphi _{3}\rangle $ is prohibited due to their identical symmetry. In
this case, we can approximate the Hamiltonian $H_{HM}(t)$ in a Hilbert space
spanned by the states $|\varphi _{1}\rangle $ and $|\varphi _{2}\rangle $

\begin{equation}
H_{HM}^{r}(t)=\left( 
\begin{array}{cc}
E_{1} & -\sqrt{2}\mu (t)/X \\ 
-\sqrt{2}\mu (t)/X & E_{2}
\end{array}
\right) ,  \eqnum{8}
\end{equation}
where $X=\sqrt{4w^{2}+E_{3}^{2}}/\sqrt{2}w$. In the case of one-photon
resonance $E_{2}-E_{1}=\omega $ and after applying the rotating-wave
approximation we obtain the expression for the probability of finding the
maximally entangled Bell state

\begin{equation}
\rho _{Bell}(t)=\frac{1}{2}(1-\cos \phi )\sin ^{2}(w\mu _{1}t/u),  \eqnum{9}
\end{equation}
in the interaction representation and weak coupling limit $u\gg w$. We can
see from Eq. (9) that the quantum state of the system at time $\tau =\pi
u/(2w\mu _{1})$ corresponds to a $\phi =\pi $ maximally entangled Bell state 
$(|RR\rangle -|LL\rangle )/\sqrt{2}$.

The result of Eq. (9) is shown in Fig. 5(c) (dotted line). Clearly, in
comparison with the exact numerical solution (solid line), our two-state
approximation describes the system evolution very well. Therefore we arrive
at the conclusion that a selective pulse of an AC field with duration $\tau
=\pi u/(2w\mu _{1})$ can be used to create and maintain a maximally
entangled Bell state in the system of two electrons in two coupled quantum
dots.

We notice from Eq. (9) that Bell-state generation time is significantly
shortened by increasing the amplitude of the AC field. This is important
because shorter Bell-state generation time is fundamental to the
experimental observation of such maximally entangled state, which is impeded
by inevitable decoherence occurred in the realistic double quantum dot
system. The decoherence is the most problematic issue pertaining to most
quantum computing processing. In the present entangled state proposal, the
decohering time depends partly on the fluctuation of the single particle
energy caused by the modification of the confining potential due to phononic
excitations. There is also a quantum electrodynamic contribution because of
coupling to the vacuum modes. In addition, impurity scattering and phonon
emission also have contributions to the decoherence. However, in principle,
their effects can be minimized by more precise fabrication technology and by
cooling the system.

\section{Summary}

In summary, we have shown how the dynamic localization and entanglement of
two interacting electrons in a coupled quantum dot system can be generated
and maintained using an AC field. The Bell-state generation time has been
calculated by an analytical approach. Exact numerical calculation confirms
these two kinds of Coulomb-involved phenomena. We hope the present study
will shed some light on the future development of the coherent control of
electrons in quantum dot systems.

\begin{center}
{\bf ACKNOWLEDGMENT}
\end{center}

This work is supported by CNSF under Grant No. 69625608 and by US-DOE.

{\huge Figure Captions}

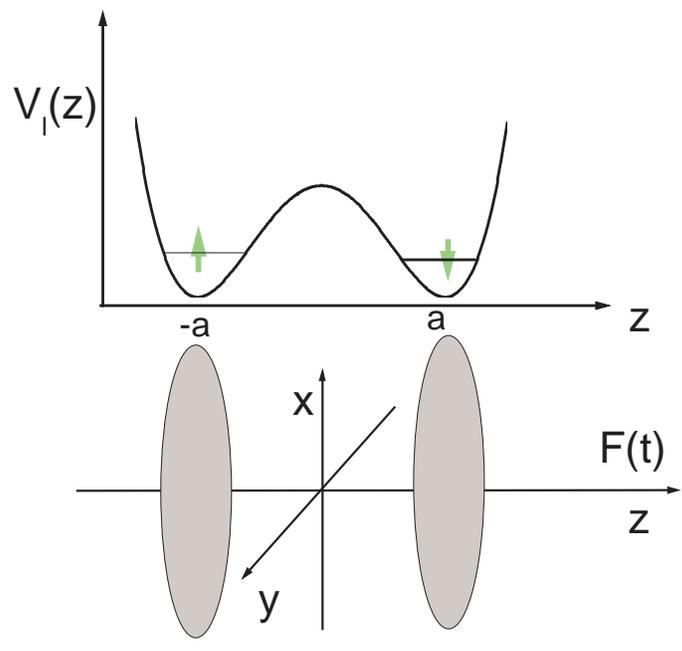
FIG. 1. Sketch{\bf \ }of two interacting electrons in a coupled quantum dot
driven by electric fields.{\bf \ }

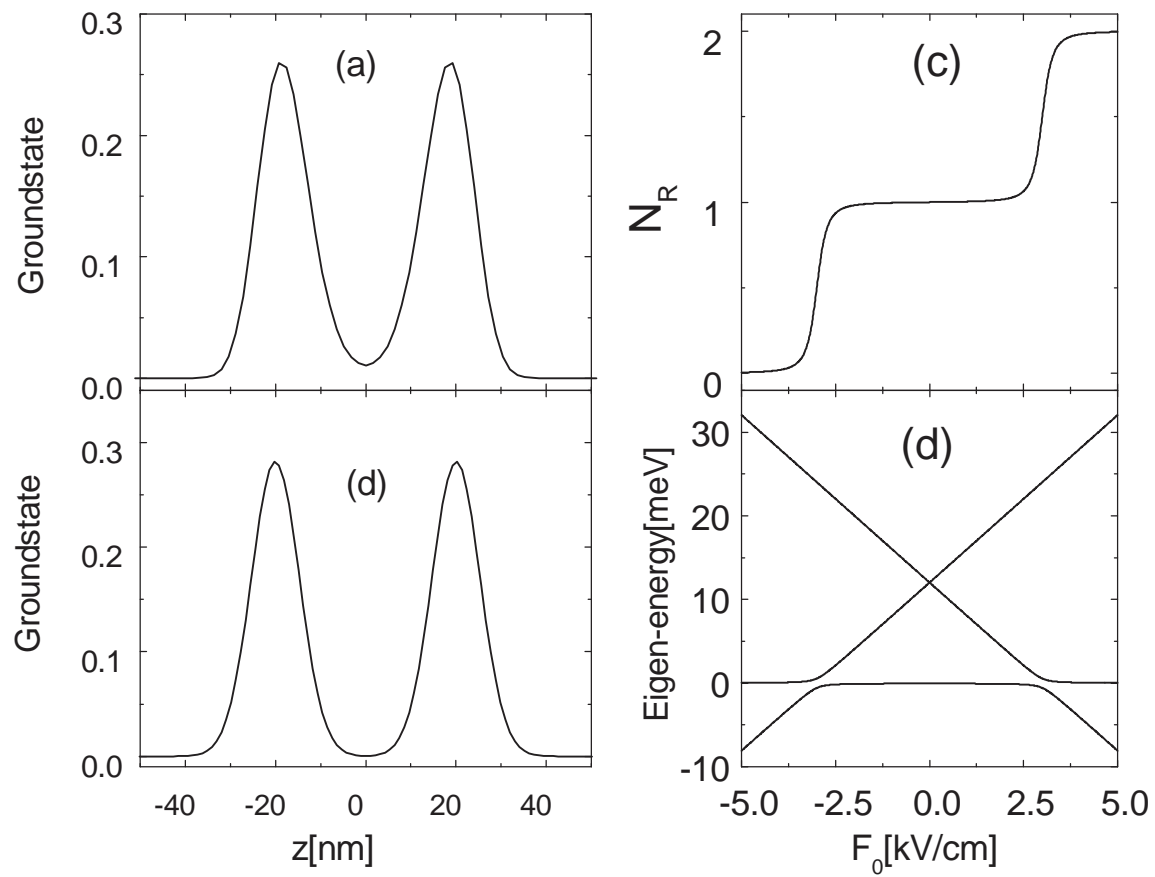
FIG. 2. (a) Two-electron ground state versus the longitudinal coordinates of
the two electrons along diagonal direction $z_{1}=-z_{2}=z$, obtained via
numerically integrating Eq. (4) with no AC field; (b) Two-electron ground
state obtained analytically via the Hund-Mulliken approximation; (c)
Electron-number distribution of the right dot in the ground state as a
function of DC electric field; (d) Low-energy spectrum versus DC field.

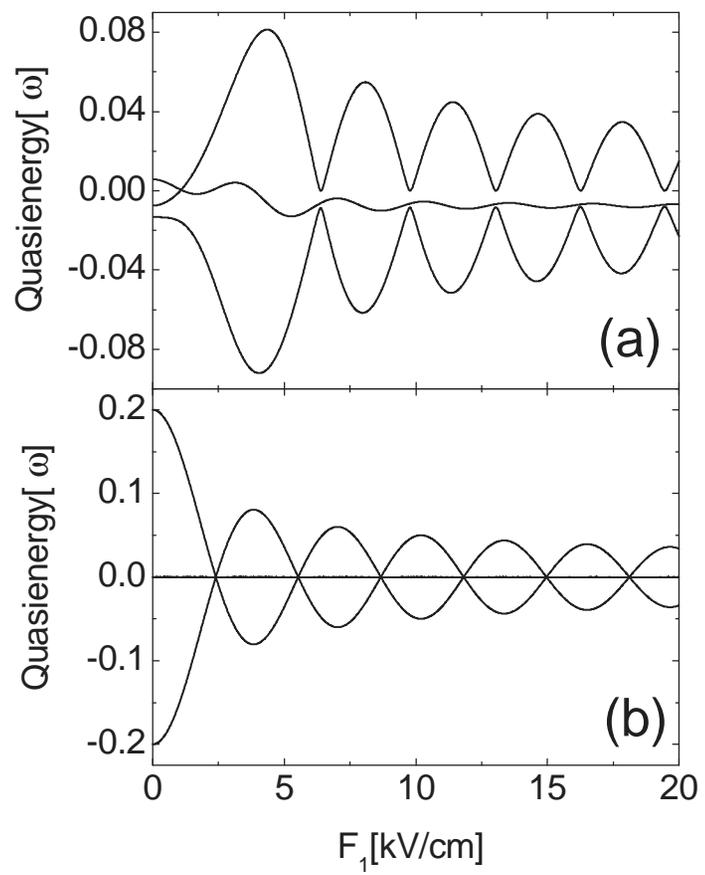
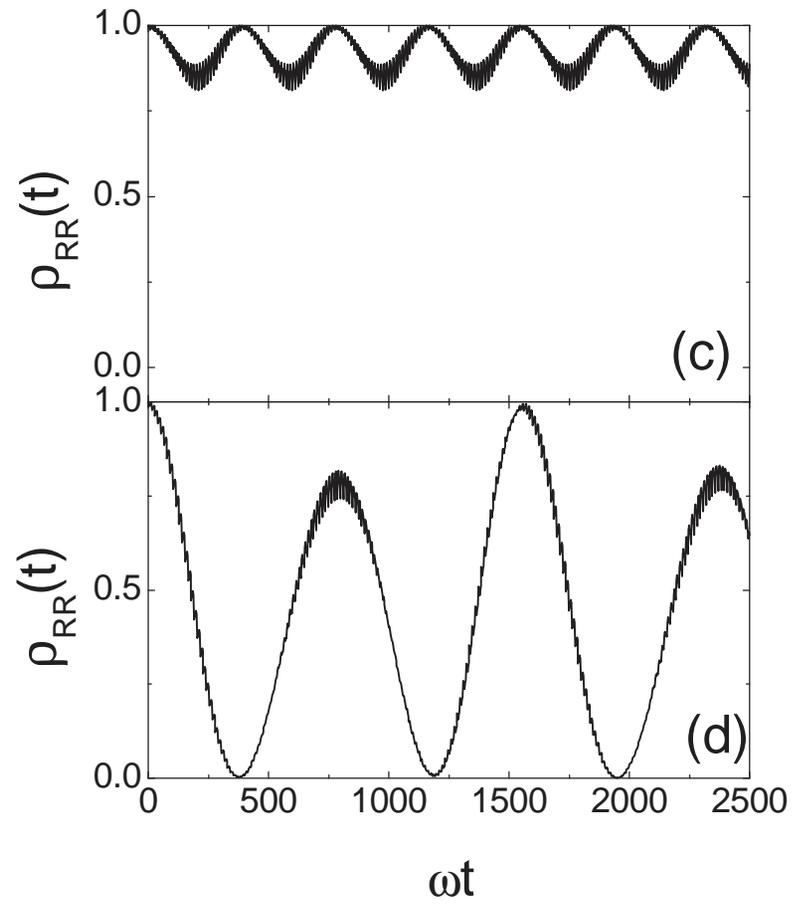
FIG. 3. (a) Quasienergy spectrum for two interacting electrons for $\omega
=4.01$meV; (b) Quasienergy spectrum with no Coulomb interaction; (c)
Probability $P_{RR}$ that two electrons occupy the same right dot for an AC
field of $F_{1}=1.078$kV/cm, which produces an exact crossing between two
quasienergies that developed from nearly-degenerate two-particle
eigenenergies; (d) $P_{RR}$ for $F_{1}=0.76$kV/cm and $\omega =6.4$meV,
corresponding to a crossing between two quasienergies, one of which
originates from unperturbed ground-state.

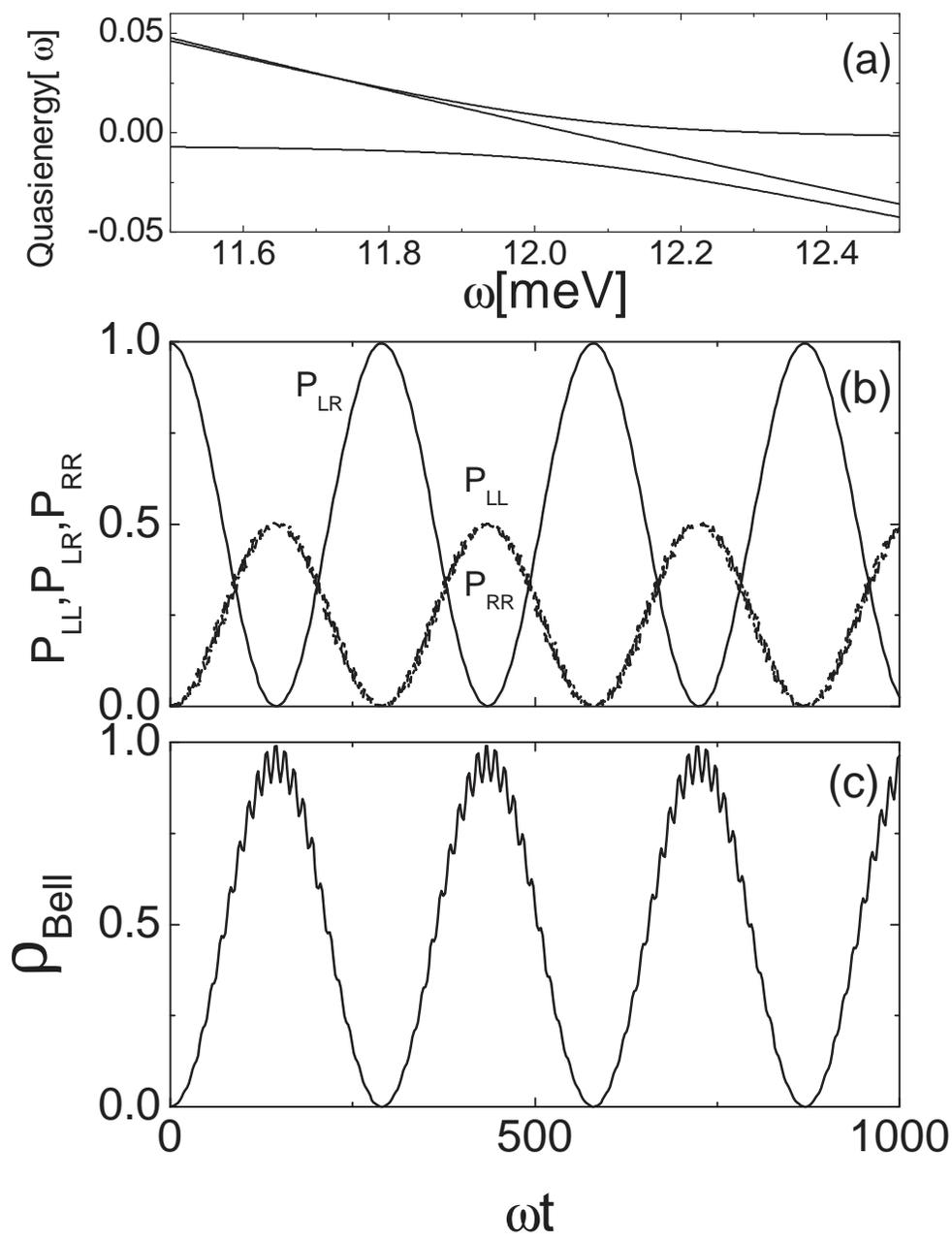
FIG. 4. (a) Floquet spectrum versus the driving frequency of an AC field
with a strength $F_{1}=1$kV/cm; (b) Probabilities that the two electrons are
in different combinations of two quantum dots, for a driving frequency $%
\omega =12$meV, corresponding to the avoided crossing shown in (a); (c)
Probability $\rho _{Bell}$ as a function of time.

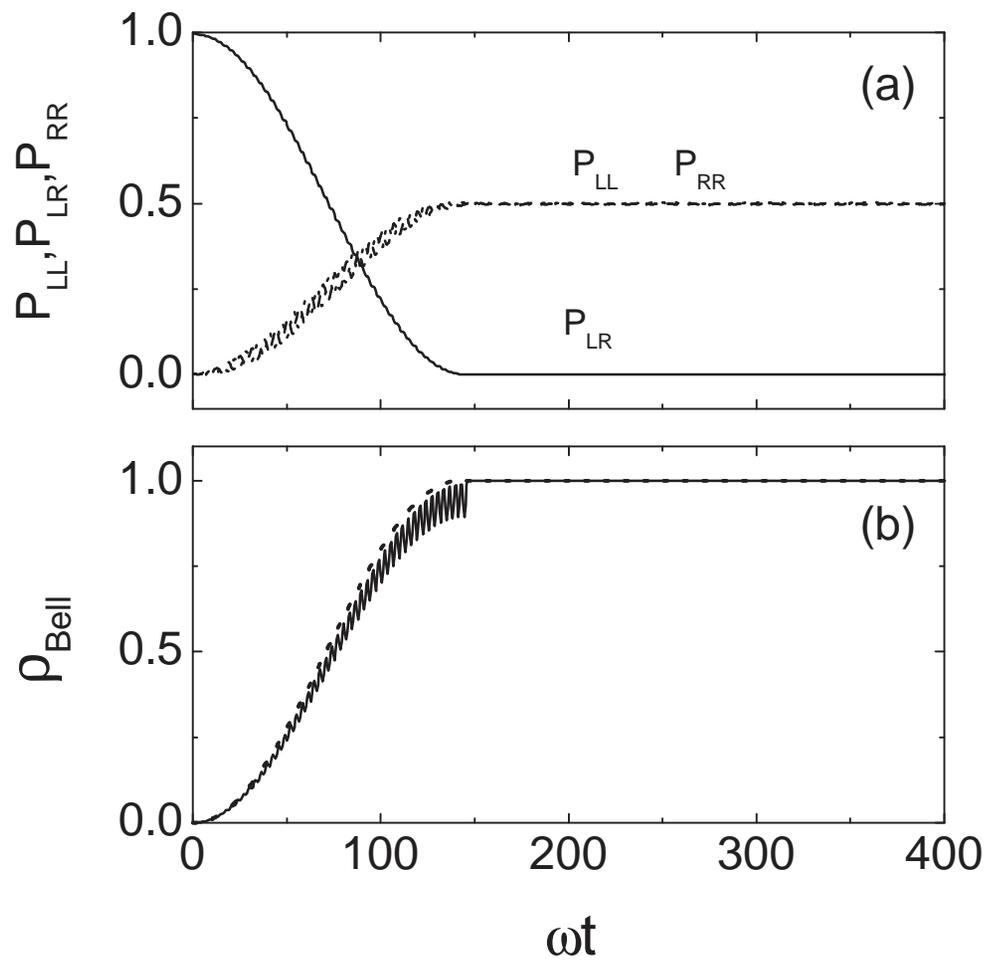
FIG. 5. (a) Probabilities that the two electrons are in different
combinations of two quantum dots, with the AC field switched off after $%
t=8.1 $ps; (b) Probability $\rho _{Bell}$ as a function of time. The
numerical and analytic results are shown with the solid and dotted lines,
respectively.

\end{document}